\documentclass[
 reprint,
superscriptaddress,
 amsmath,amssymb,
 aps,
pra,
nofootinbib,longbibliography]{revtex4-1}
\usepackage{hyperref}
\usepackage{graphicx}
\usepackage{dcolumn}
\usepackage{subcaption}
\usepackage{bm}
\usepackage{verbatim}
\usepackage{float}
\usepackage{balance}
\usepackage{natbib}
\usepackage{color}
\usepackage{hyperref}
\usepackage{graphicx}
\usepackage{dcolumn}
\usepackage{bm}

\newcommand{\bi}{\mathbf}

\begin{document}
\title{Weakened Topological Protection of the Quantum Hall Effect in a Cavity}

\author{Vasil~Rokaj}
 \email{vasil.rokaj@cfa.harvard.edu}
\affiliation{ITAMP, Center for Astrophysics $|$ Harvard $\&$ Smithsonian, Cambridge, USA}
\affiliation{Department of Physics, Harvard University, Cambridge, USA}

\author{Jie~Wang}
 \email{jiewang.phy@gmail.com}
\affiliation{Department of Physics, Harvard University, Cambridge, USA}
\affiliation{Center of Mathematical Sciences and Applications, Harvard University, Cambridge, USA}

\author{John~Sous}
\affiliation{Department of Physics, Stanford University, Stanford, USA}
\affiliation{Stanford Institute for Theoretical Physics, Stanford University, Stanford, USA}

\author{Markus~Penz}
\affiliation{Basic Research Community for Physics, Innsbruck, Austria}

\author{Michael~Ruggenthaler}
\affiliation{Max Planck Institute for the Structure and Dynamics of Matter, Hamburg, Germany}

\author{Angel~Rubio}
 \email{angel.rubio@mpsd.mpg.de}
\affiliation{Max Planck Institute for the Structure and Dynamics of Matter, Hamburg, Germany}
\affiliation{Center for Computational Quantum Physics, Flatiron Institute, New York, USA}

\begin{abstract}
 We study the quantum Hall effect in a two-dimensional homogeneous electron gas coupled to a quantum cavity field. As initially pointed out by Kohn, Galilean invariance for a homogeneous quantum Hall system implies that the electronic center of mass (CM) decouples from the electron-electron interaction, and the energy of the CM mode, also known as Kohn mode, is equal to the single particle cyclotron transition. In this work, we point out that strong light-matter hybridization between the Kohn mode and the cavity photons gives rise to collective hybrid modes between the Landau levels and the photons. We provide the exact solution for the collective Landau polaritons and we demonstrate the weakening of topological protection at zero temperature due to the existence of the lower polariton mode which is softer than the Kohn mode. This provides an intrinsic mechanism for the recently observed topological breakdown of the quantum Hall effect in a cavity [Appugliese et al., Science 375, 1030-1034 (2022)]. Importantly, our theory predicts the cavity suppression of the thermal activation gap in the quantum Hall transport. Our work paves the way for future developments in cavity control of quantum materials.
\end{abstract}

\maketitle

Interaction and topology give rise to rich exotic phases of matter, among which the integer quantum Hall (IQH) effect and the fractional quantum Hall (FQH) effect stand out~\cite{KlitzingQHE, TsuifractionalQHE, 40QHE, Laughlingfractional}. On the other side, great progress has been achieved in the manipulation of quantum materials with the use of cavity vacuum fields~\cite{ruggenthaler2017b, VidalCiutiReview, ChiralCavities, SchlawinSentefReview, kockum2019ultrastrong, KonoReview, flick2017, flick2015, RuggiReview2022, SidlerReview, Rokaj2022}. Specifically, for two-dimensional (2d) materials in magnetic fields, ultrastrong coupling of the Landau levels to the cavity field and the observation of Landau polariton quasiparticles have been achieved~\cite{ScalariScience, Keller2020, li2018, Hagenmuller2010cyclotron, rokaj2019}. Recently, modifications of the magnetotransport properties inside a cavity due to Landau polaritons were reported~\cite{paravacini2019, BartoloCiuti} and most significantly cavity modifications of the IQH transport was demonstrated~\cite{FaistCavityHall, RubioComment}. The experimental phenomena was argued to originate from a disorder-assisted cavity-mediated long-range hopping~\cite{CiutiHopping}.

In this work, given that in experiments the GaAs samples have low disorder and that the cavity field is homogeneous in the bulk of the cavity~\cite{FaistCavityHall}, we study the quantum Hall system in the homogeneous limit with vanishing disorder and we propose an alternative theory for the observed cavity modified IQH transport~\cite{FaistCavityHall}. Our theory highlights the importance of the hybridization between cavity photons and the collective Kohn mode in the quantum Hall system, and provides the exact solution for the polariton modes. In connection to the experimental findings~\cite{FaistCavityHall}, our theory draws the picture that the transport in the hybrid system is strongly influenced by the polariton states, in contrast to the standard quantum Hall transport which is purely electronic. Crucially, the low energy physics is dictated by the lower polariton mode which is softer than the cyclotron mode. The softening of the cyclotron mode signals the weakened topological protection and provides an intrinsic mechanism for the recently observed topological breakdown~\cite{FaistCavityHall}. Importantly, our theory predicts that the cavity suppresses the thermal activation gap which can be studied experimentally in the temperature dependence of the quantum Hall transport in the cavity.

\begin{figure}[H]
    \centering
    \includegraphics[width=\linewidth]{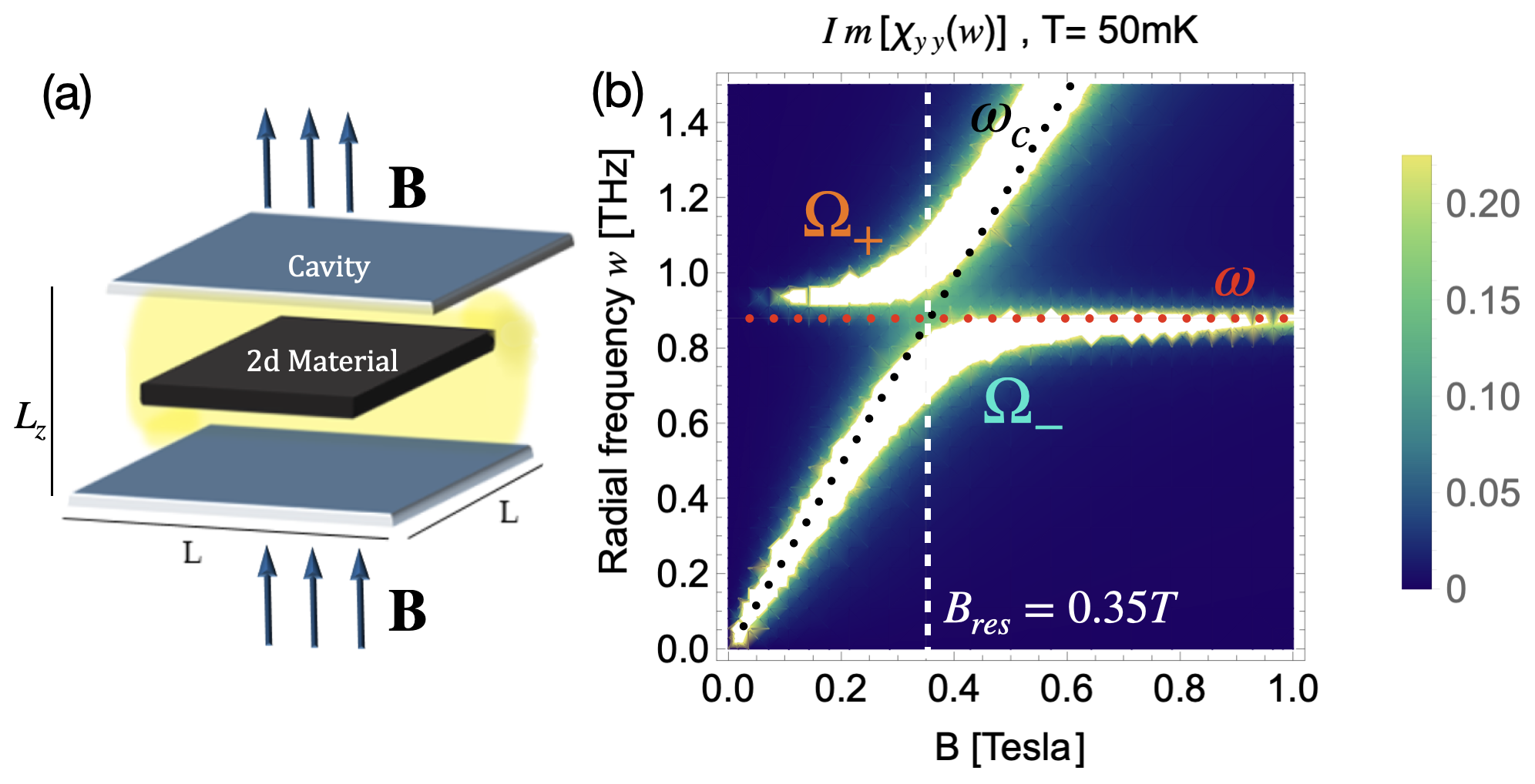}
    \caption{(a) Two-dimensional material confined in a cavity. The distance between the cavity mirrors is $L_z$. The system is placed perpendicular to a homogeneous magnetic field $\bi{B}$. (b) Imaginary part of the response function $\chi_{yy}(w)$ for IQH system in the cavity. The radial cavity frequency $\omega=2\pi\times 0.14\textrm{THz}$, the 2d electron density $n_{2d}=2\times 10^{11}\textrm{cm}^{-2}$, the effective electron mass $m=0.07m_e$ and the temperature $T=50\textrm{mK}$ are chosen according to the experiment in Ref.~\cite{FaistCavityHall}. We observe the upper $\Omega_+$ and lower $\Omega_-$ polariton, with normalized Rabi splitting $\Omega_R/\omega=0.33$. The lower polariton is softer that the cyclotron mode $\omega_c=eB/m$. This signals the weakened topological protection of the hybrid system.}
    \label{Setup}
\end{figure}

\textit{Model Hamiltonian}.---Our model considers a two-dimensional electron gas coupled to a strong magnetic field and a single-mode homogeneous cavity field, as schematically depicted in Fig.~\ref{Setup}(a). The system is described by the Pauli-Fierz Hamiltonian~\cite{spohn2004, cohen1997photons, rokaj2017}
\begin{equation}
H = \sum^{N}_{i=1}\frac{\big(\bm{\pi}_i+e\bi{A}\big)^2}{2m} + \hbar\omega\left(a^{\dagger}a + \frac{1}{2}\right) + \sum_{i<j}W(\bi{r}_i-\bi{r}_j),\label{Pauli-Fierz}
\end{equation}
where $\bm{\pi}_i=\textrm{i}\hbar\nabla_i+e\bi{A}_{\textrm{ext}}(\bi{r}_i)$ are the dynamical momenta of the electrons, and $\mathbf{A}_{\textrm{ext}}(\mathbf{r})=-\bi{e}_xBy$ describes the applied magnetic field $\bi{B}=\nabla \times \bi{A}_{\textrm{ext}}(\bi{r})=B\bi{e}_z$. The cavity field $\bi{A} = \sqrt{\frac{\hbar}{2\epsilon_0\mathcal{V}\omega}}\bi{e}_x\left(a + a^{\dagger}\right)$ is characterized by the in-plane polarization vector $\bi{e}_x$ and the photon's bare frequency $\omega$. The $\mathcal{V}$ and $\epsilon_0$ are the effective mode volume and the dielectric constant, respectively. The operators $a$ and $a^{\dagger}$ represent photonic annihilation and creation operators which satisfy bosonic commutation relations $[a,a^{\dagger}]=1$. Further, $W(\bi{r}_i-\bi{r}_j)=1/4\pi \epsilon_0 |\bi{r}_i-\bi{r}_j|$ is the Coulomb interaction between the electrons. We have parameterized the bare electron dispersion by an effective mass $m$ and assumed Galilean invariance. With Galilean invariance in a homogeneous system, the CM is decoupled from the relative motion of the electrons, regardless of the interaction strength~\cite{KohnMode}. The kinetics of the CM and its coupling to light is best described in terms of the CM coordinate $\mathbf{R}=(X,Y)=\sum^N_{i=1}\mathbf{r}_i/\sqrt{N}$ where $N$ is the total particle number. Following the derivation presented in the Supplementary Material (SM) we obtain the Hamiltonian describing the coupling of the CM to light 
\begin{equation}
H_{\textrm{cm}}=\frac{1}{2m}\left(\bm{\Pi}+e\sqrt{N}\mathbf{A}\right)^2+\hbar\omega\left(a^{\dagger}a+\frac{1}{2}\right)    
\end{equation}
where $\bm{\Pi}=\textrm{i}\hbar \nabla_{\mathbf{R}}+e\mathbf{A}_{\textrm{ext}}(\mathbf{R})$ is the dynamical momentum of the CM. It is important to mention that if we break Galilean invariance or consider a spatially inhomogeneous cavity field, the relative degrees of freedom will couple to quantum light. The CM Hamiltonian has the form of two coupled harmonic oscillators, one for the Landau level transition and one for the photons. Such a Hamiltonian is known as the Hopfield Hamiltonian which can be solved by the Hopfield transformation~\cite{Hopfield}. The Hopfield model has been employed in previous works for the description of single-particle Landau level transitions coupled to cavity photons~\cite{Hagenmuller2010cyclotron, BartoloCiuti}. Here, it shows up for the collective coupling of the electrons which emerges naturally through the CM. After the Hopfield transformation we find
\begin{equation}
   H_{\rm cm} = \hbar\Omega_+\left(b^\dag_+b_++\frac12\right) + \hbar \Omega_-\left(b^\dag_-b_-+\frac12\right)
\end{equation}
where $\{b^{\dagger}_{\pm},b_{\pm}\}$ are the creation and annihilation operators of the Landau polariton quasiparticles satisfying bosonic commutation relations $[b_l,b^{\dagger}_{l^{\prime}}]=\delta_{ll^{\prime}}$ with $l,l^{\prime}=\pm$. The details about the diagonalization of $H_{\rm{cm}}$ are given in the SM. The $\Omega_{\pm}$ are the upper and lower Landau polariton modes respectively,
\begin{eqnarray}\label{Polariton modes}
 \Omega^2_{\pm}=\frac{\omega^2+\omega^2_d+\omega^2_c}{2}\pm\sqrt{\omega^2_d\omega^2_c+\left(\frac{\omega^2+\omega^2_d-\omega^2_c}{2}\right)^2}
\end{eqnarray} 
where $\omega_d=\sqrt{e^2N/m\epsilon_0 \mathcal{V}}$ is the diamagnetic frequency originating from the $\mathbf{A}^2$ term which depends on the number of electrons $N$ and the effective mode volume $\mathcal{V}$, and $\omega_c=eB/m$ is the cyclotron frequency~\cite{Landau}. To define the polariton operators we represent $\{a,a^{\dagger}\}$ in terms of a displacement coordinate $q$ and its conjugate momentum $\partial_q$ as $a = (q+\partial_q)/\sqrt2$ with $a^{\dagger}$ obtained via conjugation~\cite{cohen1997photons, spohn2004}. The polariton operators then are written in terms of mixed coordinates as $S_{\pm}=\sqrt{\hbar/2\Omega_{\pm}}\left(b_{\pm}+b^{\dagger}_{\pm}\right)$ with
\begin{eqnarray}
    S_+ = \frac{\sqrt{m} \bar Y+q\Lambda\sqrt{\hbar/\omega}}{\sqrt{1+\Lambda^2}}\; \textrm{and}\; S_- = \frac{-q\sqrt{\hbar/\omega} +\sqrt{m} \Lambda \bar Y}{\sqrt{1+\Lambda^2}}\nonumber
\end{eqnarray}
where $\bar Y=Y+\frac{\hbar K_x}{eB}$ is the guiding center and $K_x$ is the electronic wave number in the $x$-direction. Also we introduced the parameter $\Lambda=\alpha-\sqrt{1+\alpha^2}$ with $\alpha=\left(\omega^2_c-\omega^2-\omega^2_d\right)/2\omega_d\omega_c$ which quantifies the mixing between electrons and photons.

\textit{Behavior of polaritons}.---The polariton modes $\Omega_{\pm}$ depend on the cavity frequency $\omega$, the cyclotron frequency $\omega_c$, and the number of electrons through $\omega_d$. The behavior of the polariton modes as a function of the magnetic field strength can be understood from their exact expressions Eq.~(\ref{Polariton modes}) also shown in Fig.~\ref{Setup}(b). Before the avoided crossing $\Omega_+$ follows the cavity frequency $\omega$ while $\Omega_-$ follows the cyclotron frequency $\omega_c$. After the avoided crossing the situation is inverted. On resonance $\omega=\omega_c$ the two modes are separated by the Rabi splitting $\Omega_{R}=\Omega_+-\Omega_-$ which is approximately proportional to $\omega_d$. For the geometry considered in Fig.\ref{Setup}, $\omega_d$ can be estimated through the electron density $n_{2d}$ as  $\omega_d=\sqrt{e^2N/m\epsilon_0 \mathcal{V}}=\sqrt{e^2n_{2d}\omega /\pi c m\epsilon_0}$ where we used the expression for the fundamental cavity frequency $\omega=\pi c/L_z$~\cite{rokaj2019}. Given the experimental parameters in Ref.~\cite{FaistCavityHall} for the cavity frequency $\omega=2\pi\times 0.14\textrm{THz}$, the 2d electron density $n_{2d}=2\times 10^{11}\textrm{cm}^{-2}$, and the effective electron mass $m=0.07m_e$ in GaAs, we find the normalized Rabi splitting $\Omega_R/\omega=0.33$ which is in good agreement with the experimentally observed value $\Omega_R/\omega|_{\rm exp}=0.3$~\cite{FaistCavityHall}. We note that $\omega=\omega_c$ for magnetic field strength $B_{\rm res}=0.35T$ as it is also observed experimentally~\cite{FaistCavityHall}. The normalized Rabi splitting is above $10\%$ signaling ultrastrong light-matter coupling~\cite{kockum2019ultrastrong, KonoReview}. The lower polariton is decisive for the low energy physics of the system and we will show that its behavior controls the IQH transport. Approaching the limit $\omega\rightarrow0$, the lower polariton becomes gapless reproducing the result in Refs.~\cite{rokaj2019, Rokaj2022}. In addition, $\Omega_-$ decreases as a function of the light-matter coupling strength, controlled via $\omega_d$, i.e., $\Omega_- < \omega_c$ when $\omega_d > 0$. In what follows, we discuss the implications of the polariton states for the quantum Hall transport at zero and finite temperature.

\textit{Fragility of topological protection against polariton lifetimes and ultrastrong light-matter coupling.}---A clean or weakly disordered quantum Hall system at zero temperature, as long as it is gapped, is expected to be topologically protected~\cite{Thouless}. However, the softening of the cyclotron mode, due to the lower polariton, indicates that the topological protection of the system is weakened. Due to the gap reduction, the transport of the system can be more easily affected by disorder, which leads to a finite lifetime for the polariton quasiparticles. The polariton lifetimes will be included phenomenologically and we will see that their effect combined with ultrastrong coupling enables the breakdown of topological protection~\cite{FaistCavityHall, RubioComment}.

The gauge-invariant current operator for homogeneous fields solely depends on the CM dynamical momentum and the cavity field~\cite{Landau, Rokaj2022} $\mathbf{J}=-\frac{e\sqrt{N}}{m}\left(\mathbf{\Pi}+e\sqrt{N}\mathbf{A}\right)$. Due to this property and the separability of $H_{\textrm{cm}}$ from the electronic correlations we can compute the transport of the system by focusing only on the states of $H_{\rm cm}$. At $T=0$ the system is in the polariton vacuum $|\Psi_{\textrm{gs}}\rangle=|0_+\rangle|0_-\rangle$ which is annihilated by both polariton operators $b_{\pm}$. Given this state, we employ the standard Kubo formalism~\cite{kubo} for the computation of the current correlators $\chi_{ab}(t)=-\textrm{i}\Theta(t) \langle \Psi_{\rm gs}|[J_{a}(t),J_{b}]|\Psi_{\rm gs}\rangle/\hbar$ in the time domain which we transform to the frequency domain in order to obtain the optical conductivities~\cite{kubo} $\sigma_{ab}(w)=\frac{\textrm{i}}{w+\textrm{i}\delta}\left(\frac{e^2n_{2d}}{m}\delta_{ab}+\frac{\chi_{ab}(w)}{A}\right)$ where $A$ and $n_{2d}=N/A$ are the area and the electron density of the 2d material respectively, $\delta$ is the broadening parameter, and $\delta_{ab}$ the Kronecker delta with $a,b \in \{x,y\}$. The optical conductivities $\sigma_{ab}(w)$ are given in the frequency domain in terms of the frequency $w$. The full details for the transport computations are provided in the SM. The poles of the response functions $\chi_{ab}(w)$ identify the optical responses of the system and its excitations. As we show in Fig.~\ref{Setup}(b) the optical excitations correspond to Landau polariton modes, which have been observed in a multitude of experiments ~\cite{FaistCavityHall, Keller2020, li2018, Langebeyondunity}. Note that in Fig.~\ref{Setup}(b) we use the parameters reported in Ref.~\cite{FaistCavityHall} which we described previously. In addition, using the Kubo formula we find the Hall and longitudinal DC ($w=0$) conductivities  
\begin{eqnarray}
      \sigma_{xy}&=&\frac{e^2\nu}{h(1+\Lambda^2)}\left[\frac{\Lambda(\Lambda+\eta)}{\Omega^2_-/\omega^2_c+\delta^2/\omega^2_c}+\frac{1-\eta\Lambda}{\Omega^2_+/\omega^2_c+\delta^2/\omega^2_c}\right]\nonumber\\
\sigma_{yy}&=&\sigma_{D}\left[1-\frac{1}{1+\Lambda^2}\left(\frac{\Omega^2_+}{\Omega^2_+ +\delta^2} +\frac{\Lambda^2 \Omega^2_-}{\Omega^2_-+\delta^2}\right)\right]
\end{eqnarray}
where $\eta=\omega_d/\omega_c$. Note that $\sigma_{D}=e^2n_{2d}/m\delta$ is the Drude DC conductivity, and that in $\sigma_{xy}$ we introduced the Landau level filling factor $\nu=n_{2d}h/eB$~\cite{Peierls, Tong}. For $\delta \rightarrow 0$ we find the Hall conductance perfectly quantized $\sigma_{xy}=e^2\nu/h$, consistent with the Thouless flux insertion argument~\cite{Thouless}. In the last step we used two properties of the mixing parameter $
 1- \eta \Lambda = \Omega^2_+/\omega^2_c\;\;\textrm{and}\;\; \Lambda(\Omega^2_-/\omega^2_c-1)=\eta
$ which are deduced from the definition of $\Lambda$. 

The polariton lifetimes are responsible for the broadening in the transmission spectra observed experimentally~\cite{FaistCavityHall, Keller2020, li2018, Langebeyondunity}. The total lifetime is a result of several mechanisms: scattering by impurities, radiative decay~\cite{cohen1997photons}, coupling to phonons, as well as to the substrate. Here, we phenomenologically model the polariton lifetime as $\tau=1/\delta$ by keeping a finite broadening $\delta$ which enables to model the experimental optical spectra as for example in Fig.~\ref{Setup}(b).

\begin{figure}[H]
    \centering
    \includegraphics[width=
\linewidth]{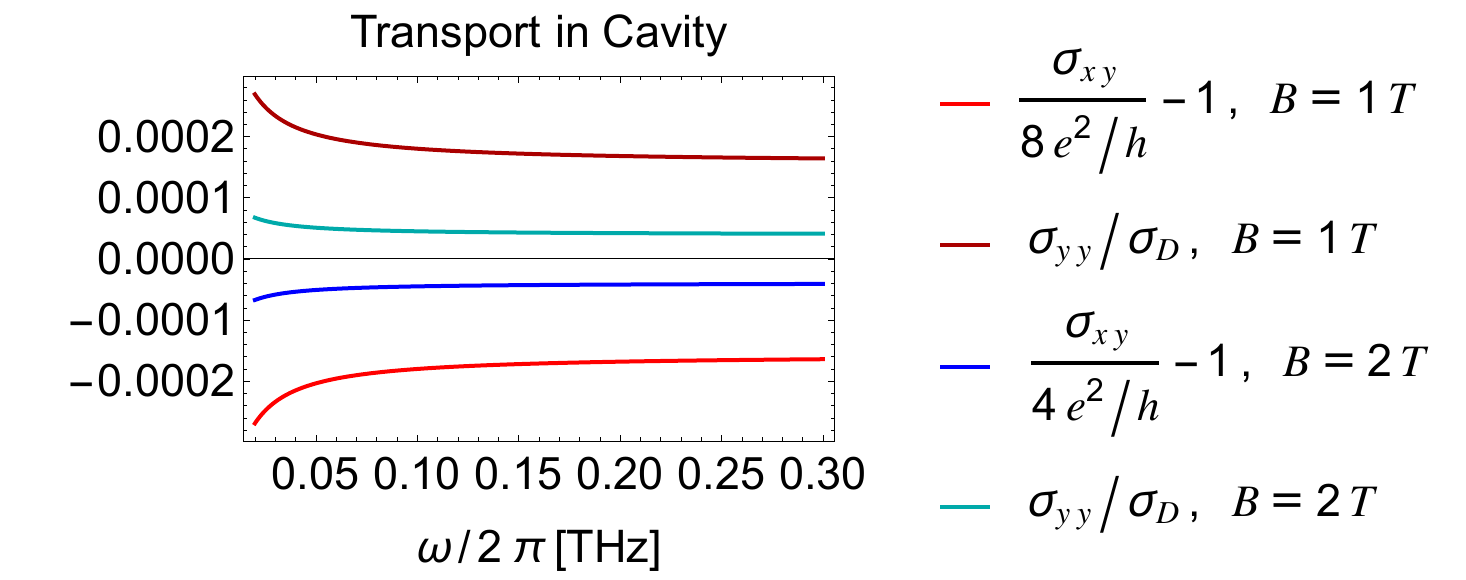}
    \caption{Quantum Hall transport in a cavity at $T=0$ with a finite broadening $\delta=2\pi \times 5\times 10^{-3}\textrm{THz}$ for two different values of magnetic field strength $B=1T$ and $B=2T$ which correspond to the filling factors $\nu=8$ and $\nu=4$ respectively. The 2d electron density is $n_{2d}=2\times 10^{11}\textrm{cm}^{-2}$ as in Ref.~\cite{FaistCavityHall}. The Hall and the longitudinal conductivities deviate from the topologically expected values. For the smaller value of the magnetic field (higher filling factor) the deviations of the IQH transport due to the cavity are enhanced.}
    \label{Lifetime}
\end{figure}
Motivated by the experiments in Refs.~\cite{li2018, paravacini2019,FaistCavityHall} we choose $\delta=2\pi \times 5\times10^{-3}\textrm{THz}$ and in Fig.\ref{Lifetime} we plot $\sigma_{xy}$ and $\sigma_{yy}$ under ultrastrong light-matter coupling for different values of the magnetic field strength, corresponding to different filling factors. In Fig.\ref{Lifetime} we see that $\sigma_{xy}/(\nu e^2/h)$ deviates from unity and $\sigma_{yy}$ deviates from zero. Both phenomena signal the breakdown of topological protection. For $B=2T$ we observe that the cavity effects are suppressed in comparison to $B=1T$. This is physically expected as for larger magnetic fields the vacuum field fluctuations become a small perturbation to the system. The deviations from the expected values occur off-resonance, for a small cavity frequency, because in this regime the lower polariton gap $\Omega_-$ is significantly reduced (see Fig.~\ref{Setup}(b)). This relates to the fact that for fixed electron density and small $\omega$ the normalized Rabi splitting $\Omega_R/\omega$ is enhanced. Thus, it is the interplay between the ultrastrong light-matter coupling and the finite polariton lifetime that causes the effects on transport. This intuitive physical picture is in agreement with the observed breakdown of topological protection in Ref~\cite{FaistCavityHall}, and the disorder-assisted cavity-mediated hopping mechanism~\cite{CiutiHopping}. 

The above analysis is consistent with the result in the long-wavelength limit $\omega \rightarrow 0$ and $\delta=0$~\cite{RokajButterfly2021}. The Hall conductivity for $\delta=0$ is quantized for all $\omega >0$ (finite gap) but drops to $e^2\nu/h/(1 + \eta ^2)$ for $\omega \rightarrow 0$ (gapless)~\cite{RokajButterfly2021}. At this point the canonical transformation to the polariton basis becomes singular. In this sense, the broadening $\delta$ regularizes the long-wavelength limit result.

\textit{Cavity suppression of the thermal activation gap}.---Finite temperature transport properties are also strongly influenced by coupling the electrons to the cavity. This can be understood from the formula for the thermal behavior of the longitudinal transport $
    \sigma_{yy}(T)/\sigma_{yy}(T = 0) \approx \exp\left(-\beta\Delta\right)$
where $\Delta$ is the activation gap of the system and $\beta=1/kT$. For the hybrid system, $\Delta = \Omega_-$. Thereby for the IQH effect, the coupling to the cavity generally speaking reduces the activation gap from $\omega_c$ to $\Omega_-$ and makes the Hall transport easier to be modified by temperature. 
\begin{figure}[H]
    \centering
    \includegraphics[width=
\linewidth]{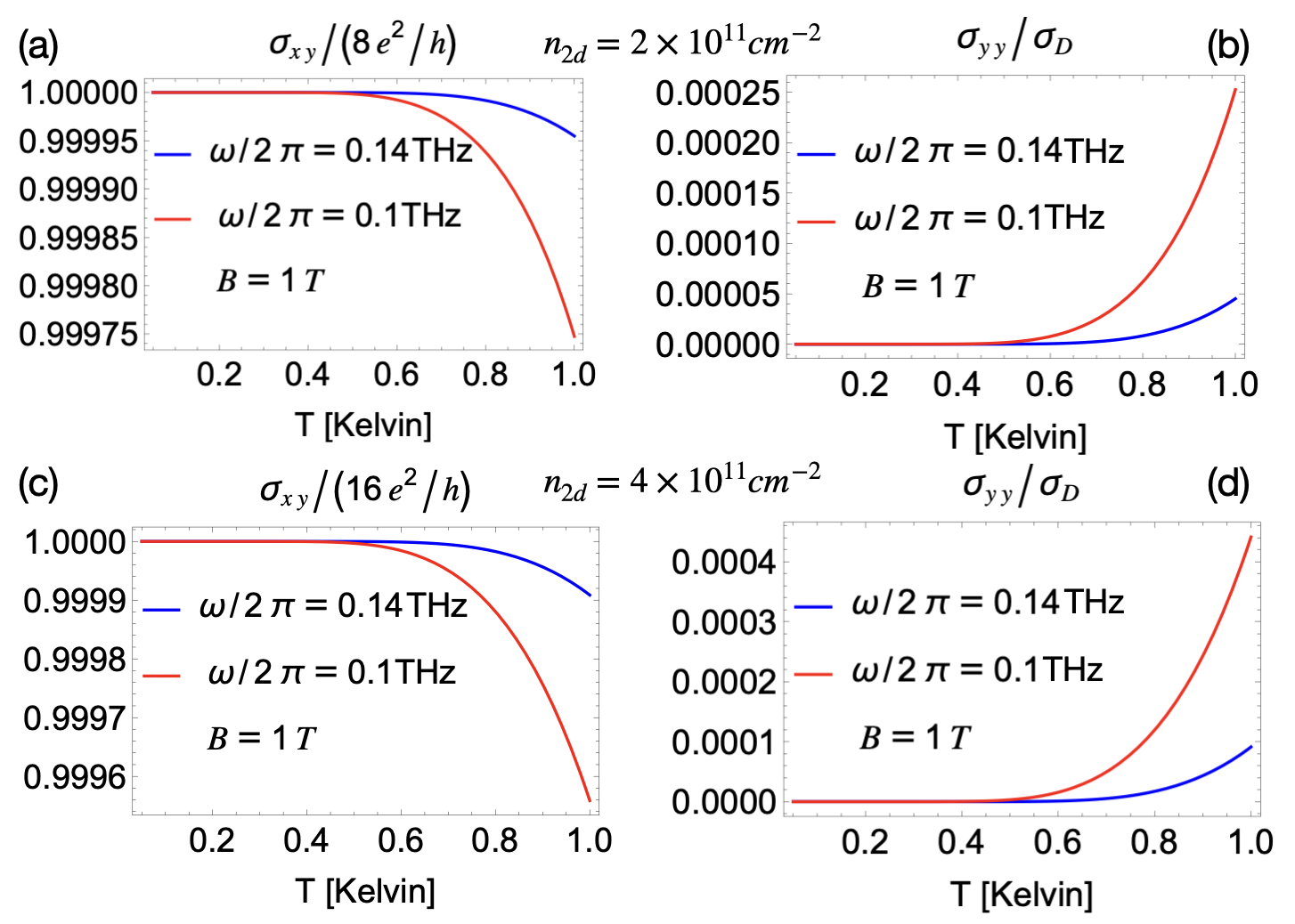}
    \caption{Low temperature transport at magnetic field strength $B=1T$. In (a) and (b) $n_{2d}=2\times 10^{11}\textrm{cm}^{-2}$ while in (c) and (d) $n_{2d}=4\times 10^{11}\textrm{cm}^{-2}$. The light-matter coupling strongly affects the quantum Hall transport. For the smaller cavity frequency $\omega=2\pi \times 0.1$THz and the larger electron density the deviation from the topologically protected values maximizes. This relates to the behavior of $\Omega_-$ which controls the thermal activation in the system. The broadening parameter is chosen very small $\delta=2\pi\times 10^{-4}$THz to avoid influencing transport, but guarantee numerical convergence.}
    \label{Temperature}
\end{figure} The quantitative description of the thermal activation gap is obtained from the finite temperature Kubo formula~\cite{kubo}, through $\chi_{\alpha\beta}(w)$ which is the retarded current correlation function,
\begin{equation}
    \chi_{ab}(w)=\sum_{M,Q}\frac{e^{-\beta E_{M}}-e^{-\beta E_Q}}{\mathcal{Z}}   \frac{\langle \Psi_M|J_{a}|\Psi_Q\rangle\langle \Psi_{Q}|J_{b}|\Psi_{M}\rangle}{w+(E_M-E_Q)/\hbar +\textrm{i}\delta}
\end{equation}
where $|\Psi_M\rangle,|\Psi_Q\rangle$ are the many-body states  with eigenenergies $E_M,E_Q$ respectively, and $\mathcal{Z}$ is the partition function $\mathcal{Z} = \sum_M \exp(-\beta E_M)$. The details of the temperature dependent transport are given in the SM. For the temperature dependent computations presented in Fig.~\ref{Temperature} we use two set of parameters for the cavity frequency $\omega=2\pi \times\{0.14,0.1\}$THz and the 2d electron density $n_{2d}=\{2,4\}\times 10^{11}\textrm{cm}^{-2}$. The first ones correspond to the parameters reported in Ref.~\cite{FaistCavityHall} and the second are used for comparison. The magnetic field strength is $B=1T$. Figure~\ref{Temperature} demonstrates that the finite temperature transport of the IQH system can be modified by the cavity. From the behavior of both conductivities it is evident that the dependence of transport on temperature is enhanced for the lower cavity frequency $\omega=2\pi \times 0.1$THz. This is directly connected to the gap reduction in the system as $\Omega_-$ takes a smaller value for a smaller $\omega$. Additionally, we observe that the temperature effect is also enhanced by the electron density by comparing Figs.\ref{Temperature}~(a) and (b) to Figs.\ref{Temperature}~(c) and (d). This is to be expected since the electron density controls the Rabi splitting $\Omega_R$. It is important to mention that in comparison to Fig.~\ref{Lifetime}, the broadening parameter in  Fig.~\ref{Temperature} is one order of magnitude smaller $\delta=2\pi \times 10^{-4}\textrm{THz}$ such that effect from the polariton lifetime becomes negligible. In the SM we provide the finite temperature transport for the parameters used in Fig.~\ref{Lifetime}. The low temperature transport is consistent with the $T=0$ results in Fig.~\ref{Lifetime}. 

\textit{Connections to Experiments and Future Directions.}---The above analysis suggests that the activation gap of the hybrid system is strongly suppressed by coupling to cavity modes. Importantly, our model enables the theoretical estimate of the activation gap and direct comparison to experiment. 

Additionally, we discuss the difference reported experimentally between the odd and the even plateaus~\cite{FaistCavityHall}. The odd plateaus in the IQH effect are due to the Zeeman gap. In the experiment~\cite{FaistCavityHall} the Zeeman gap is $20\%$ of the cyclotron gap. Thus, we can think of the Zeeman gap effectively as a cyclotron gap with an effective magnetic field strength reduced by the factor $1/5$ as compared to the actual magnetic field, i.e., $\Delta_{\rm Zeeman}=\omega_c/5=eB_{\rm eff}/m$ where $B_{\rm eff}=B/5$. Under this assumption we compute the deviations of IQH transport in the cavity for the odd plateau. The deviation of the longitudinal transport from zero for $\omega=2\pi \times 0.14$THz, at $T=0$ and $B=1T$ is $\sigma_{yy}/\sigma_{D}|_{B}=1.7\times 10^{-4}$ (see also Fig.~\ref{Lifetime}) while for the respective odd plateau with $B_{\rm eff}=0.2T$ is one order of magnitude larger, $\sigma_{yy}/\sigma_{D}|_{B_{\rm eff}}=4.3\times 10^{-3}$. The Hall conductivities behave similarly as it can be understood from Fig.~\ref{Lifetime}. This analysis shows that odd plateaus are much more vulnerable to the cavity than the even ones. A more rigorous treatment of this effect requires the inclusion of the spin degrees of freedom. This is an interesting problem for future investigation.

Further, we comment on the FQH effect. In samples with low disorder, the activation gap of the FQH effect is given by the many-body gap, closely related to the magneto-roton energy~\cite{Magnetoroton}, which we assume to be smaller than $\Omega_-$ and therefore protected from cavity effects. This picture is consistent with the experimental observations that FQH plateaus are relatively immune to the cavity~\cite{FaistCavityHall}. From this analysis we anticipate that the FQH effect can be modified at low temperature when $\Omega_-$ becomes softer than the many-body gap.

To summarize, using a Galilean invariant quantum Hall model coupled to a homogeneous single-mode cavity field, we provide the exact solution for the polariton states and discuss their experimental implications for quantum Hall transport in cavities. The lower polariton is softer than the cyclotron mode and leads to the weakening of topological protection. This provides an intrinsic mechanism for the recently observed breakdown of the topological protection of the IQH effect due to cavity vacuum fluctuations~\cite{FaistCavityHall}. Having understood analytically the homogeneous setting, our work paves the way for future investigations going beyond this limit, such that the interaction between the polaritons and the electron correlations comes into play. In this setting the interplay between polaritons and anyons is an interesting future research question, with potential applications to quantum computing~\cite{AnyonsDasSarma}. It is important to mention that in Ref.~\cite{FaistCavityHall} in the edges of the sample the cavity field is not perfectly homogeneous. Despite this fact the bulk polariton modes are observed in the transmission spectrum. This proves the robustness of the polaritons against the field inhomogeneities. Nevertheless, it is an interesting future direction to study the influence of the inhomogeneities of the cavity field to edge modes in disordered samples. The inclusion of impurities will be important for a more precise understanding of transport. Incorporating leakage and the multimode structure of the cavity will enable a more realistic description of transport phenomena in general electromagnetic environments~\cite{svendsenRealCavities}. Finally, we highlight that the quantum Hall system has played a crucial role in redefining units in terms of constants of nature~\cite{vonKlitzingUnits}. Thus, the cavity induced phenomena could potentially have implications for metrology as pointed out in Ref.~\cite{FaistCavityHall}.

\begin{acknowledgments}
We would like to thank J. Faist, F. Appugliese, J. Enkner, and L. Graziotto and  for fruitful discussions. V. R. acknowledges support from the NSF through a grant for ITAMP at Harvard University. J. S. acknowledges support from the Gordon and Betty Moore Foundation’s EPiQS Initiative through Grant GBMF8686 at Stanford University. This work is also supported from the Cluster of Excellence `CUI: Advanced Imaging of Matter'- EXC 2056 - project ID 390715994, SFB-925 ``Light induced dynamics and control of correlated quantum systems" – project 170620586  and Grupos Consolidados (IT1453-22). We acknowledge support from the Max Planck-New York City Center for Non-Equilibrium Quantum Phenomena. J. W. acknowledges the support from Flatiron institute where this project was initialized. The Flatiron Institute is a division of the Simons Foundation.
\end{acknowledgments}

\bibliography{cavity_QED.bib}

\clearpage
\pagebreak

\onecolumngrid
\widetext

\begin{center}
\textbf{ SUPPLEMENTARY MATERIAL} \\
\end{center}

\section{Pauli-Fierz Hamiltonian in the Center of Mass Frame}\label{CM Hamiltonian}

In this section we would like to give the details about the transformation of the Pauli-Fierz Hamiltonian in the center of mass (CM) and relative distances frame. The Hamiltonian of our system is 
\begin{eqnarray}
H&=&\frac{1}{2m}\sum^{N}_{i=1}\big(\bm{\pi}_i+e\bi{A}\big)^2+\sum\limits^{N}_{i< l}W\left(|\mathbf{r}_i-\mathbf{r}_l|\right)+\hbar\omega\left(a^{\dagger}a+\frac{1}{2}\right).
\end{eqnarray}
where $\bm{\pi}_i=\textrm{i}\hbar\nabla_i+e\mathbf{A}_{\textrm{ext}}(\mathbf{r}_i)$ are the dynamical momenta, and $\mathbf{A}_{\textrm{ext}}(\mathbf{r})=-\bi{e}_xB y$ describes the homogeneous magnetic field $\bi{B}=\nabla \times \bi{A}_{\textrm{ext}}(\bi{r})=B\bi{e}_z$. The cavity field $\bi{A} = \sqrt{\frac{\hbar}{2\epsilon_0\mathcal{V}\omega}}\bi{e}_x\left(a + a^{\dagger}\right)$ is characterized by the in-plane polarization vector $\bi{e}_x$ and the photon's bare frequency $\omega$. Further, $W(\bi{r}_i-\bi{r}_j)=1/4\pi \epsilon_0 |\bi{r}_i-\bi{r}_j|$ is the Coulomb interaction between the electrons. For mathematical convenience we utilize a symmetric definition with respect to $\sqrt{N}$ for the coordinates in the CM frame as in Ref.~\cite{BuschSol}
\begin{equation}
   \mathbf{R}=\frac{1}{\sqrt{N}}\sum^N_{i=1}\mathbf{r}_i \;\; \textrm{and}\;\; \widetilde{\mathbf{r}}_{j}=\frac{\mathbf{r}_1-\mathbf{r}_j}{\sqrt{N}}\;\;\textrm{with}\;\;j>1.
\end{equation}
The original electronic coordinates in terms of the new ones $\{\mathbf{R},\widetilde{\mathbf{r}}_j\}$ are
\begin{eqnarray}
   \mathbf{r}_1=\frac{1}{\sqrt{N}}\left(\mathbf{R}+\sum^N_{j=2}\widetilde{\mathbf{r}}_j\right)\;\textrm{and}\; 
   \mathbf{r}_j=\frac{1}{\sqrt{N}}\left(\mathbf{R}+\sum^N_{j=2}\widetilde{\mathbf{r}}_j\right) -\sqrt{N}\widetilde{\mathbf{r}}_j\;\;\textrm{with}\;\;j>1.\nonumber
\end{eqnarray}
The momenta of the electrons in the new coordinate system are $\nabla_1=\left(\nabla_{\mathbf{R}}+\sum^N_{j=2}\widetilde{\nabla}_j\right)/\sqrt{N}$ and $
\nabla_{j}= \left(\nabla_{\mathbf{R}}-\widetilde{\nabla}_j\right)/\sqrt{N}\;\textrm{with}\;j>1.$
From these expression we can find the form of the electronic kinetic terms in the new frame
\begin{eqnarray}\label{momenta}
&&\sum^N_{i=1}\nabla^2_i=\nabla^2_{\mathbf{R}} +\frac{1}{N}\sum^N_{j=2}\widetilde{\nabla}^2_{j} +\frac{1}{N}\sum^N_{j,k=2}\widetilde{\nabla}_{j}\cdot \widetilde{\nabla}_{k}\;\; \textrm{and}\;\; \sum^N_{i=1}\nabla_i=\sqrt{N} \nabla_{\mathbf{R}}.
\end{eqnarray}
The interaction term between the cavity field and the electrons takes the form
\begin{equation}\label{CM Bfield}
    \mathbf{A}\cdot \sum^N_{i=1}\textrm{i}\hbar\nabla_i+e\mathbf{A}_{\textrm{ext}}(\mathbf{r}_i)=\sqrt{N}\mathbf{A}\cdot\left(\textrm{i}\hbar \nabla_{\mathbf{R}}+e\mathbf{A}_{\textrm{ext}}(\mathbf{R})\right)
\end{equation}

 To complete our analysis we also give the expression for the purely electronic terms in the new frame. For the quadrature of the external magnetic field we have
\begin{eqnarray}
\sum^N_{i=1}\mathbf{A}^2_{\textrm{ext}}(\mathbf{r}_i)=\mathbf{A}^2_{\textrm{ext}}(\mathbf{R})+N\sum^N_{j=2}\mathbf{A}^2_{\textrm{ext}}(\widetilde{\mathbf{r}}_j)-\left[\sum^N_{j=2}\mathbf{A}_{\textrm{ext}}(\widetilde{\mathbf{r}}_j)\right]^2\nonumber\\
\end{eqnarray}
and for the bilinear term between the magnetic field and the momenta we have
\begin{eqnarray}
\sum^N_{i=1}\mathbf{A}_{\textrm{ext}}(\mathbf{r}_i)\cdot\nabla_i=\mathbf{A}_{\textrm{ext}} (\mathbf{R})\cdot \nabla_{\mathbf{R}}+\sum^{N}_{j=2}\mathbf{A}_{\textrm{ext}}(\widetilde{\mathbf{r}}_j)\cdot \widetilde{\nabla}_{j}.\nonumber\\
\end{eqnarray}
Finally, we give the expression for the interaction term $W$ between the electrons.
\begin{eqnarray}
 \sum\limits^{N}_{i< l} W\left(|\mathbf{r}_i-\mathbf{r}_l|\right)=\sum^N_{1<l}W(\sqrt{N}|\widetilde{\mathbf{r}}_l|)+\sum^N_{2\leq i<l}W\left(\sqrt{N}|\widetilde{\mathbf{r}}_i-\widetilde{\mathbf{r}}_l|\right)\nonumber
\end{eqnarray}
Adding together all the different terms we find that the expression of the Hamiltonian in the new frame is the sum of two parts: (i) the center of mass part $H_{\textrm{com}}$ which is coupled to the quantized field $\mathbf{A}$ and (ii) the relative distances $H_{\textrm{rel}}$ which does not couple to the cavity field, $H=H_{\textrm{cm}}+H_{\textrm{rel}}$
where each part looks as
\begin{eqnarray}
 H_{\textrm{cm}}&=&\frac{1}{2m}\left(\textrm{i}\hbar\nabla_{\mathbf{R}}+e\mathbf{A}_{\textrm{ext}}(\mathbf{R})+e\sqrt{N}\mathbf{A}\right)^2+\hbar\omega\left(a^{\dagger}a+\frac{1}{2}\right),\nonumber\\
 H_{\textrm{rel}}&=&\frac{1}{2m}\sum^N_{j=2}\left(\frac{\textrm{i}\hbar}{\sqrt{N}}\widetilde{\nabla}_{j}+e\sqrt{N}\mathbf{A}_{\textrm{ext}}(\widetilde{\mathbf{r}}_j)\right)^2 -\frac{\hbar^2}{2mN}\sum^N_{j,l=2}\widetilde{\nabla}_{j}\cdot\widetilde{\nabla}_{l}-\frac{e^2}{2m}\left(\sum^N_{j=2}\mathbf{A}_{\textrm{ext}}(\widetilde{\mathbf{r}}_j)\right)^2\nonumber\\
 &+&\sum^N_{1<l}W(\sqrt{N}|\widetilde{\mathbf{r}}_l|)+\sum^N_{2\leq i<l}W\left(\sqrt{N}|\widetilde{\mathbf{r}}_i-\widetilde{\mathbf{r}}_l|\right).
\end{eqnarray}
Lastly, it is important to demonstrate that the center of mass and relative distances degrees of freedom are independent by checking the commutation relations between their coordinates and momenta. Using the chain rule we have for the derivatives of the coordinates in the center of mass frame
\begin{eqnarray}
\nabla_{\mathbf{R}}=\sum^N_{i=1}\frac{\partial \mathbf{r}_i}{\partial \mathbf{R} }\nabla_i=\frac{1}{\sqrt{N}}\sum^N_{i=1}\nabla_i, \;\; \textrm{and}\;\;\widetilde{\nabla}_j=\sum^N_{i=1}\frac{\partial \mathbf{r}_i}{\partial \widetilde{\mathbf{r}}_j}\nabla_i=\frac{1}{\sqrt{N}}\sum^N_{i=1}\nabla_i -\sqrt{N}\nabla_j\;\textrm{with}\;j>1.\nonumber
\end{eqnarray}
From the above expressions it is clear that the momenta in the new coordinate frame commute 
\begin{eqnarray}
\left[\nabla_{\mathbf{R}},\widetilde{\nabla}_j\right]=0.
\end{eqnarray}
The next property to check is the commutation relations between the momenta and the coordinates.
\begin{eqnarray}
 \left[\nabla_{\mathbf{R}},\widetilde{\mathbf{r}}_j \right]&=&\frac{1}{N}\left[\sum^N_{i=1}\nabla_i,\mathbf{r}_1-\mathbf{r}_j\right]=0,\;\;\; j> 1,\\
    \left[\widetilde{\nabla}_j,\mathbf{R} \right]&=&\left[\frac{1}{\sqrt{N}}\sum^N_{i=1}\nabla_i-\sqrt{N}\nabla_j,\frac{1}{\sqrt{N}}\sum^N_{l=1}\mathbf{r}_l\right]=\nonumber\\
    &=&\frac{1}{N}\sum^N_{i,l=1}\left[\nabla_i,\mathbf{r}_l\right]-\sum^N_{l=1}\left[\nabla_j,\mathbf{r}_l\right]=\frac{1}{N}\sum^N_{i,l=1}\delta_{il}-\sum^N_{l=1}\delta_{jl}=\frac{N}{N}-1=0.
\end{eqnarray}
We would also like to mention that the separation between the CM and the relative distances holds true also for an arbitrary amount of photon modes as long as the cavity field is considered to be homogeneous as it was shown in~\cite{rokaj2022cavity}.

\section{Exact Solution of the CM Hamiltonian and Landau Polaritons}\label{Polariton Solution}
Having demonstrated that only the CM of the electronic system couples to the cavity we will show that $H_{\textrm{cm}}$ can be solved analytically. Let us see how this can be done. To proceed we expand the covariant kinetic term
\begin{equation}
    H_{\textrm{cm}}=\frac{\bm{\Pi}^2}{2m}+\frac{e\sqrt{N}}{m}\mathbf{A}\cdot\bm{\Pi}+ \underbrace{\frac{e^2N\mathbf{A}^2}{2m}+\hbar\omega\left(a^{\dagger}a +\frac{1}{2}\right)}_{H_p}
\end{equation}
For the description of the photon operators we will introduce the displacement coordinate $q$ and its conjugate momentum $\partial_q$ as $a=\frac{1}{\sqrt{2}}\left(q+\partial/\partial q\right)$ and $a^{\dagger}$ defined by conjugation~\cite{spohn2004, GriffithsQM}. The part $H_p$ can be brought to diagonal form by the scaling transformation on the photonic displacement coordinate
\begin{equation}
    u=q\sqrt{\frac{\widetilde{\omega}}{\omega}}\;\; \textrm{where} \;\;\widetilde{\omega}=\sqrt{\omega^2+\omega^2_d}
\end{equation}
with $\omega_d=\sqrt{e^2N/\epsilon_0m \mathcal{V}}$ is the diamagnetic frequency depending on the electron density in the effective mode volume. After this transformation the CM Hamiltonian is
\begin{eqnarray}
   H_{\textrm{cm}}=\frac{\bm{\Pi}^2}{2m}+\frac{e\sqrt{N}}{m}\mathbf{A}\cdot \bm{\Pi}+\frac{\hbar\widetilde{\omega}}{2}\left(-\frac{\partial^2}{\partial u^2}+u^2\right),
\end{eqnarray}
where the quantized field is now 
\begin{equation}
    \mathbf{A}=\sqrt{\frac{\hbar}{\epsilon_0V\widetilde{\omega}}}\mathbf{e}_xu.
\end{equation}
In the Landau gauge the Hamiltonian has translational invariance along the $X$ coordinate which implies that the eigenfunctions in $X$ are plane waves $e^{\textrm{i}K_xX}$. We apply $H_{\textrm{cm}}$ on the plane wave and we have
\begin{eqnarray}
   H_{\textrm{cm}}=-\frac{\hbar^2}{2m}\frac{\partial^2}{\partial Y^2}+\frac{m\omega^2_c}{2}\left(Y+\frac{\hbar K_x}{eB}\right)^2-geB u\left(Y+\frac{\hbar K_x}{eB}\right)+\frac{\hbar\widetilde{\omega}}{2}\left(-\frac{\partial^2}{\partial u^2}+u^2\right)\nonumber
\end{eqnarray}
where we also introduced the coupling constant $g=\omega_d\sqrt{\hbar/m\widetilde{\omega}}$. As a next step we define the coordinate
\begin{equation}\label{Y bar}
    \bar Y=Y+\frac{\hbar K_x}{eB}
\end{equation}
and the Hamiltonian simplifies further
\begin{eqnarray}
   H_{\textrm{cm}}=-\frac{\hbar^2}{2m}\frac{\partial^2}{\partial \bar Y^2}+\frac{m\omega^2_c}{2}\bar Y^2-geB u \bar Y+\frac{\hbar\widetilde{\omega}}{2}\left[-\frac{\partial^2}{\partial u^2}+u^2\right]\nonumber
\end{eqnarray}
The Hamiltonian consists of two coupled harmonic oscillators. It is convenient to perform another scaling transformation on $\bar Y$ and $u$ 
\begin{equation}\label{Vpm}
    V_-=-u\sqrt{\frac{\hbar}{\widetilde{\omega}}}\;\; \textrm{and}\;\; V_+=\sqrt{m}\bar Y.
\end{equation}
such that we have both harmonic oscillators in the form of having mass equal to 1. The Hamiltonian then becomes
\begin{equation}
     H_{\textrm{cm}}=-\frac{\hbar^2}{2}\sum_{l=\pm}\frac{\partial^2}{\partial V^2_{l}}+\frac{1}{2}\sum_{l,j=\pm}W_{lj}V_{ l}V_{ j}.
\end{equation}
The matrix $W$ 
\begin{eqnarray}
 W=\left(\begin{tabular}{ c c c c }
		$\omega^2_c$ & $\omega_d\omega_c $ \\
	    $\omega_d\omega_c $ &$\widetilde{\omega}^2$ 
	\end{tabular}\right)
\end{eqnarray}
is real and symmetric, and as a consequence can be diagonalized by the orthogonal matrix $O$~\cite{faisal1987}, 
\begin{eqnarray}\label{Omatrix}
 &&O=\left(\begin{tabular}{ c c c c }
	    $\frac{1}{\sqrt{1+\Lambda^2}} $ &$\frac{\Lambda}{\sqrt{1+\Lambda^2}}$ \\$-\frac{\Lambda}{\sqrt{1+\Lambda^2}}$ & $\frac{1}{\sqrt{1+\Lambda^2}} $ 
	\end{tabular}\right)\;\;\; \textrm{where}\;\; \Lambda=\alpha-\sqrt{1+\alpha^2} \;\;\; \textrm{and}\;\;\; \alpha=\frac{\omega^2_c-\widetilde{\omega}^2}{2\omega_d\omega_c}.\nonumber
\end{eqnarray}
The eigenvalues of the matrix $W$ give the new normal modes of the interacting light-matter system. We find them to be
\begin{eqnarray}
 \Omega^2_{\pm}&=&\frac{1}{2}\left(\widetilde{\omega}^2+\omega^2_c\pm\sqrt{4\omega^2_d\omega^2_c+(\widetilde{\omega}^2-\omega^2_c)^2}\right).
\end{eqnarray} 
The Hamiltonian after the orthogonal transformation takes the canonical form
\begin{eqnarray}
 H_{\textrm{cm}}=-\frac{\hbar^2}{2}\sum_{l=\pm}\frac{\partial^2}{\partial S^2_{ l}}+\frac{1}{2}\sum_{l=\pm}\Omega^2_{l}S^2_{ l}.
\end{eqnarray}
The new coordinates $S_{l}$ and conjugate momenta $\partial_{S_{l}}$ are related to the old ones $\{V_{l},\partial_{V_{l}}\}$ through the orthogonal matrix $O$,
\begin{equation}\label{SandV}
    S_{l}=\sum_{j=\pm}O_{jl}V_{j} \; \textrm{and}\; \frac{\partial}{\partial S_{ l}}=\sum_{j=\pm}O_{jl}\frac{\partial}{\partial V_{ j}}.
\end{equation}
Due to the fact that the matrix $O$ is orthogonal the canonical commutation relations are satisfied which implies that we have two independent harmonic oscillators~\cite{faisal1987}. Thus, the eigenfunctions of the interacting system are Hermite functions $\Phi$ of the coordinates $S_{+}$ and $S_{-}$. The full set of eigenfunctions of the system is 
\begin{equation}
    \Psi_{K_x,n_+,n_-}(X,S_{+},S_{-})=e^{\textrm{i}K_xX}\Phi_{n_+}(S_{ +})\Phi_{n_-}(S_{ -})
\end{equation}
with eigenspectrum
\begin{equation}
    E_{n_+,n_-}=\hbar\Omega_+\left(n_+ +\frac{1}{2}\right)+\hbar\Omega_-\left(n_- +\frac{1}{2}\right).
\end{equation}
The frequencies $\Omega_+$ (upper) and $\Omega_-$ (lower) are the two collective Landau polariton modes of the quantum Hall system in the cavity. For completeness, we note that the solution of the polaritons for the CM can be equivalently written in terms of annihilation $b_{\pm}$ and creation $b^{\dagger}_{\pm}$ operators for the polariton quasiparticles. In this representation $H_{\rm cm}$ is written as 
\begin{eqnarray}
    H_{\rm cm }=\hbar \Omega_+\left(b^{\dagger}_{+}b_++\frac{1}{2}\right)+ \hbar \Omega_-\left(b^{\dagger}_{-}b_-+\frac{1}{2}\right)
\end{eqnarray}
with the polariton operators defined $b_{\pm}=S_{\pm}\sqrt{\frac{\Omega_{\pm}}{2}} +\sqrt{\frac{1}{2\Omega_{\pm}}}\partial_{S_{\pm}}$~\cite{GriffithsQM}. It is worth to notice that in the limit $\omega \rightarrow 0$ the lower polariton frequency goes to zero, $\Omega_-\rightarrow 0$, which means that the system becomes gapless. In this limit the canonical transformation from the electron and photon basis $V_{\pm}$ to the polariton basis $S_{\pm}$ becomes singular.

\section{Finite Temperature Transport}\label{Transport}

In this section we present the general formalism employed for the finite temperature transport of the light-matter system. As we already showed the Hamiltonian of our system can be written as a sum of a CM and relative part $H=H_{\rm cm}+H_{\rm rel}$. To proceed we assume that the eigenstates of $H_{\rm cm}$ are $|\Phi_n\rangle$ and the eigenstates of $H_{\rm rel}$ are $|F_{I}\rangle$ such that it holds
\begin{eqnarray}
H_{\rm cm}|\Phi_{n}\rangle=E_n|\Phi_n\rangle \;\; \textrm{and}\;\; H_{\rm rel}|F_{I}\rangle=E_{I}|F_{I}\rangle
\end{eqnarray}
Then, the eigenstates of the full Hamiltonian $H$ are
\begin{equation}
    |\Psi_{nI}\rangle=|\Phi_n\rangle\otimes |F_{I}\rangle,
\end{equation}
and the full eigenspectrum is $E_{nI}=E_{n}+E_{I}$. The Kubo formula for the optical conductivity of the system is~\cite{kubo, ALLEN2006165}
\begin{eqnarray}
 \sigma_{ab}(w)=\frac{\textrm{i}}{w+\textrm{i}\delta}\left(\frac{e^2n_e}{m}\delta_{ab}+\frac{\chi_{ab}(w)}{A}\right)\;\;\; \delta \rightarrow 0^+
\end{eqnarray}
where $a,b=x,y,z$. The first term in the optical conductivity is the Drude term, while the second term is the current-current correlator in the frequency domain, which is defined as the Fourier transform of current-current correlator in the time domain
\begin{eqnarray}
     \chi_{ab}(t)=\frac{-\textrm{i}\Theta(t)}{\hbar }\langle[J_{a}(t),J_{b}]\rangle,
 \end{eqnarray}
with the current operators considered in the interaction picture $\mathbf{J}(t)=e^{\textrm{i}Ht/\hbar}\mathbf{J}e^{-\textrm{i}Ht/\hbar}$~\cite{kubo}. In the canonical ensemble the expectation value of an operator $\mathcal{O}$ is defined as~\cite{ALLEN2006165}
\begin{eqnarray}
    \langle \mathcal{O}\rangle=Tr\{\rho \mathcal{O}\}=\frac{1}{\mathcal{Z}}\sum_{n,I} \langle \Psi_{nI}|e^{-\beta H\mathcal{O}}|\Psi_{nI}\rangle
\end{eqnarray}
where the partition function is $\mathcal{Z}=\sum_{n,I}e^{-\beta E_n}e^{-\beta E_{I}}$. We will use these formulas now for the computation of the current correlation functions. The current response can be splitted into two parts 
\begin{eqnarray}\label{JJResponse}
     \chi_{ab}(t)=\frac{-\textrm{i}\Theta(t)}{\hbar }\left(\langle J_a(t) J_b\rangle -\langle J_b J_a(t)\rangle \right).
 \end{eqnarray}
Let us compute first the first term $\langle J_a(t) J_b\rangle$. We use the expression for the canonical ensemble and for the current operator in the interaction picture and we have
\begin{eqnarray}
    \langle J_a(t) J_b\rangle&=&\frac{1}{\mathcal{Z}}\sum_{n,I}e^{-\beta E_{nI}} \langle \Psi_{nI}|e^{\textrm{i}Ht/\hbar} J_ae^{-\textrm{i} Ht/\hbar} J_b|\Psi_{nI}\rangle\nonumber\\
    &=&\frac{1}{\mathcal{Z}}\sum_{n,I}e^{-\beta E_{nI}} e^{\textrm{i}tE_{nI}/\hbar} \langle \Psi_{nI}| J_ae^{-\textrm{i} Ht/\hbar} J_b|\Psi_{nI}\rangle.\nonumber\\
\end{eqnarray}
We introduce the identity $\mathbb{I}=\sum_{m,J}|\Psi_{mJ}\rangle\langle \Psi_{mJ}|$ in the above expression 
\begin{eqnarray}
  \langle J_a(t) J_b\rangle&=& \frac{1}{\mathcal{Z}}\sum_{n,m,J,I}e^{-\beta E_{nI}} e^{\textrm{i}tE_{nI}/\hbar} \langle \Psi_{nI}| J_ae^{-\textrm{i} Ht/\hbar} |\Psi_{mJ}\rangle\langle \Psi_{mJ}| J_b|\Psi_{nI}\rangle \nonumber \\
  &=& \frac{1}{\mathcal{Z}}\sum_{n,m,J,I}e^{-\beta E_{nI}} e^{\textrm{i}t(E_{nI}-E_{mJ})/\hbar}  \langle \Psi_{nI}| J_a|\Psi_{mJ}\rangle\langle \Psi_{mJ}| J_b|\Psi_{nI}\rangle 
\end{eqnarray}

\subsection{Current in the CM frame}

Since we work in the CM frame in order to proceed we need examine how the current operator looks in the CM frame. The expression for the current operator can be obtained by computing the velocity operator of the electrons through the Heisenberg equation of motion~\cite{Landau}
\begin{equation}
 \mathbf{v}_i=\frac{d \mathbf{r}_i}{dt}=\frac{\textrm{i}}{\hbar} [H,\mathbf{r}_i]=\frac{1}{m}\left(-\textrm{i}\hbar \nabla_i-e\mathbf{A}_{\textrm{ext}}(\mathbf{r}_i)-e\mathbf{A}\right).
\end{equation}
Then, the full gauge-invariant current operator is~\cite{Landau}
\begin{eqnarray}\label{Current Operator}
    \bi{J}=e\sum^N_{i=1}\mathbf{v}_i=-\frac{\textrm{i}e\hbar}{m_{\textrm{e}}}\sum^N_{j=1}\nabla_j-\frac{e^2N}{m_{\textrm{e}}}\bi{A}-\frac{e^2}{m_{\textrm{e}}}\sum^N_{i=1}\bi{A}_{\textrm{ext}}(\mathbf{r}_i).\nonumber\\
\end{eqnarray}
We to go to the CM and relative distances frame and we utilize the expressions derived in Appendix~\ref{CM Hamiltonian} for all the relevant operators and we find for current operator
\begin{equation}\label{Current COM}
    \mathbf{J}=\sqrt{N}\left[-\frac{\textrm{i}e\hbar}{m}\nabla_{\mathbf{R}}-\frac{e^2}{m}\sqrt{N}\mathbf{A}-\frac{e^2}{m}\mathbf{A}_{\textrm{ext}}(\mathbf{R})\right]\equiv \mathbf{J}_{\rm cm}.
\end{equation} 
The above result shows that the total current in the system is equal essentially to current of the CM and depends only on CM related operators. This property has the following important implication
\begin{equation}
    \langle \Psi_{nI}|\mathbf{J}|\Psi_{mJ}\rangle=\delta_{IJ} \langle \Phi_n|\mathbf{J}|\Phi_m\rangle
\end{equation}
using the above the expression for the current correlator simplifies 
\begin{eqnarray}
  \langle J_a(t) J_b\rangle=\frac{1}{\mathcal{Z}}\sum_{n,m,I}e^{-\beta E_{nI}} e^{\textrm{i}t(E_{n}-E_{m})/\hbar}  \langle \Phi_n|J_{a}|\Phi_m\rangle\langle \Phi_{m}|J_{b}|\Phi_{n}\rangle\nonumber\\ 
\end{eqnarray}
We note to obtain the above we used that $E_{nI}-E_{mI}=E_{n}-E_{m}$. To complete the computation we need to multiply $\langle J_a(t) J_b\rangle$ with $\frac{-\textrm{i}\Theta(t)}{\hbar }$ and Fourier transform into the frequency space

    \begin{eqnarray}
    \frac{-\textrm{i}\Theta(t)}{\hbar } \langle J_a(t) J_b\rangle & \longrightarrow & \frac{1}{\mathcal{Z}}\sum_{n,m,I}e^{-\beta E_{nI}}   \frac{\langle \Phi_n|J_{a}|\Phi_m\rangle\langle \Phi_{m}|J_{b}|\Phi_{n}\rangle}{w+(E_n-E_m)/\hbar +\textrm{i}\delta}=\frac{\sum_{I}e^{-\beta E_{I}}}{\sum_{I}e^{-\beta E_{I}}\sum_{k}e^{-\beta E_k}} \sum_{n,m,I}e^{-\beta E_{n}}   \frac{\langle \Phi_n|J_{a}|\Phi_m\rangle\langle \Phi_{m}|J_{b}|\Phi_{n}\rangle}{w+(E_n-E_m)/\hbar +\textrm{i}\delta}\nonumber \\
    &&=\frac{1}{\sum_{k}e^{-\beta E_k}} \sum_{n,m}e^{-\beta E_{n}}   \frac{\langle \Phi_n|J_{a}|\Phi_m\rangle\langle \Phi_{m}|J_{b}|\Phi_{n}\rangle}{w+(E_n-E_m)/\hbar +\textrm{i}\delta} \;\; \textrm{with}\;\; \delta \rightarrow 0^+.
\end{eqnarray}

Following exactly the same procedure for the second term in Eq.(\ref{JJResponse}) $\frac{\textrm{i}\Theta(t)}{\hbar } \langle \mathbf{J}\mathbf{J}(t)\rangle$ we find the the expression for the current-current response function
\begin{equation}
    \chi_{ab}(w)=\frac{1}{\sum_{l}e^{-\beta E_l}} \sum_{n,m}\left(e^{-\beta E_{n}}-e^{-\beta E_m}\right)   \frac{\langle \Phi_n|J_{a}|\Phi_m\rangle\langle \Phi_{m}|J_{b}|\Phi_{n}\rangle}{w+(E_n-E_m)/\hbar +\textrm{i}\delta} \;\; \textrm{with}\;\; \delta \rightarrow 0^+.
\end{equation}
From the above expression we see that current response function solely depends on the CM eigenstates and the CM eigenenergies. This is a consequence of homogeneity which implies the separability of the full Hamiltonian into CM and relative parts.

\subsection{Application to Landau Polaritons}

Having derived the general formula for the current response function of a homogeneous system, we will now apply to the Landau polaritons. For the polaritons  we have the CM eigenstates $e^{\textrm{i}K_xX}\phi_{n_+}(S_+)\phi_{n_-}(S_-)\equiv |K_xn_+n_-\rangle$ and the eigenergies $E_{n_+n_-}=\hbar\Omega_+\left(n_++\frac{1}{2}\right) + \hbar\Omega_-\left(n_-+\frac{1}{2}\right)$. Consequently the response functions take the form
\begin{equation}
    \chi_{ab}(w)=\sum_{n_+,n_-,m_+,m_-,\\ K_x,K^{\prime}_x}\frac{e^{-\beta E_{n_+n_-}}-e^{-\beta E_{m_+m_-}}}{\mathcal{Z}_{\rm cm}}   \frac{\langle n_+n_-K^{\prime}_x|J_{a}|K_xm_+m_-\rangle\langle K_xm_+m_-|J_{b}|K^{\prime}_xn_+n_-\rangle}{w+(E_{n_+n_-}-E_{m_+m_-})/\hbar +\textrm{i}\delta} 
\end{equation}
where $\mathcal{Z}_{\rm cm}=\sum_{n_+,n_-,K_x}e^{-\beta E_{n_+n_-}}$ is the CM partition function. To proceed further we need the expressions for the current operators in the polaritonic basis. The $x$ and $y$ components of the current operator in terms of the polaritonic coordinates ${S_\pm}$ are
\begin{eqnarray}
 J_x&=&\frac{e^2\sqrt{N}B}{m^{3/2}_e}\left[\frac{\sqrt{m}}{eB}\left(-\textrm{i}\hbar\nabla_X-\hbar K_x\right)+\frac{S_+(1-\eta\Lambda)+S_-(\Lambda+\eta)}{\sqrt{1+\Lambda^2}}\right]\nonumber\\
 J_y&=&-\frac{\textrm{i}e\hbar}{m}\sum^N_{j=1}\partial_{y_j}=\frac{-\textrm{i}e\hbar}{\sqrt{m}}\sqrt{\frac{N}{1+\Lambda^2}}\left[\partial_{S_+}+\Lambda\partial_{S_-}\right].
\end{eqnarray}
Moreover, the current operators can be written using the polaritonic annihilation and creation operators as follows
\begin{eqnarray}
    J_x&=&\frac{e^2\sqrt{N}B}{m^{3/2}_e}\sqrt{\frac{\hbar}{2(1+\Lambda^2)}}\left[\frac{\sqrt{m}}{eB}\left(-\textrm{i}\hbar\nabla_X-\hbar K_x\right)+\frac{\Lambda+\eta}{\sqrt{\Omega_-}}\left(b^{\dagger}_-+b_-\right)+\frac{1-\eta\Lambda}{\sqrt{\Omega_+}}\left(b^{\dagger}_++ b_+\right)\right]\\
J_y&=&-\textrm{i}e\sqrt{\frac{ \hbar N}{2m(1+\Lambda^2)}}\left[\sqrt{\Omega_+}\left(b_+-b^{\dagger}_+\right)+\Lambda\sqrt{\Omega_-}\left(b_--b^{\dagger}_-\right)\right]
\end{eqnarray}
From the above we can obtain the matrix representation of the current operator on the polariton basis
   \begin{eqnarray}
\langle n_+n_-K^{\prime}_x|J_x|K_xm_+m_-\rangle&=&\frac{e^2\sqrt{N}B}{m^{3/2}_e}\sqrt{\frac{\hbar}{2(1+\Lambda^2)}}\Bigg[\frac{\Lambda+\eta}{\sqrt{\Omega_-}}\delta_{n_+m_+}\left(\sqrt{m_-+1}\delta_{n_-,m_-+1}+\sqrt{m_-}\delta_{n_-,m_--1}\right)\nonumber\\
    &+&\frac{1-\eta\Lambda}{\sqrt{\Omega_+}}\delta_{n_-m_-}\left(\sqrt{m_+}\delta_{n_+,m_+-1}+\sqrt{m_++1}\delta_{n_+,m_++1}\right)\Bigg]\delta_{K^{\prime}_xK_x}\\
\langle n_+n_-K^{\prime}_x|J_y|K_xm_+m_-\rangle&=&-\textrm{i}e\sqrt{\frac{ \hbar N}{2m(1+\Lambda^2)}}\Big[\sqrt{\Omega_+}\delta_{n_-m_-}\left(\sqrt{m_+}\delta_{n_+,m_+-1}-\sqrt{m_++1}\delta_{n_+,m_++1}\right)\nonumber\\
    &+&\Lambda\sqrt{\Omega_-}\delta_{n_+m_+}\left(\sqrt{m_-}\delta_{n_-,m_--1}-\sqrt{m_-+1}\delta_{n_-,m_-+1}\right)\Big]\delta_{K^{\prime}_xK_x}
\end{eqnarray} 
The current operators are diagonal with respect to the plane-wave states $e^{\textrm{i}K_xX}$ and consequently the current response functions simplifies to
    \begin{equation}
    \chi_{ab}(w)=\sum_{n_+,n_-,m_+,m_-}\frac{e^{-\beta E_{n_+n_-}}-e^{-\beta E_{m_+m_-}}}{\sum_{l_+,l_-} e^{-\beta E_{l_+l_-}}}   \frac{\langle n_+n_-|J_{a}|m_+m_-\rangle\langle m_+m_-|J_{b}|n_+n_-\rangle}{w+(E_{n_+n_-}-E_{m_+m_-})/\hbar +\textrm{i}\delta} 
\end{equation}
With the use of the above formula for current response functions the temperature dependent transport of the polariton system can be obtained. The corresponding results are shown in main text of the manuscript. 

\begin{figure}[H]
     \centering
     \begin{subfigure}[b]{0.35\textwidth}
         \centering
        \includegraphics[width=\columnwidth]{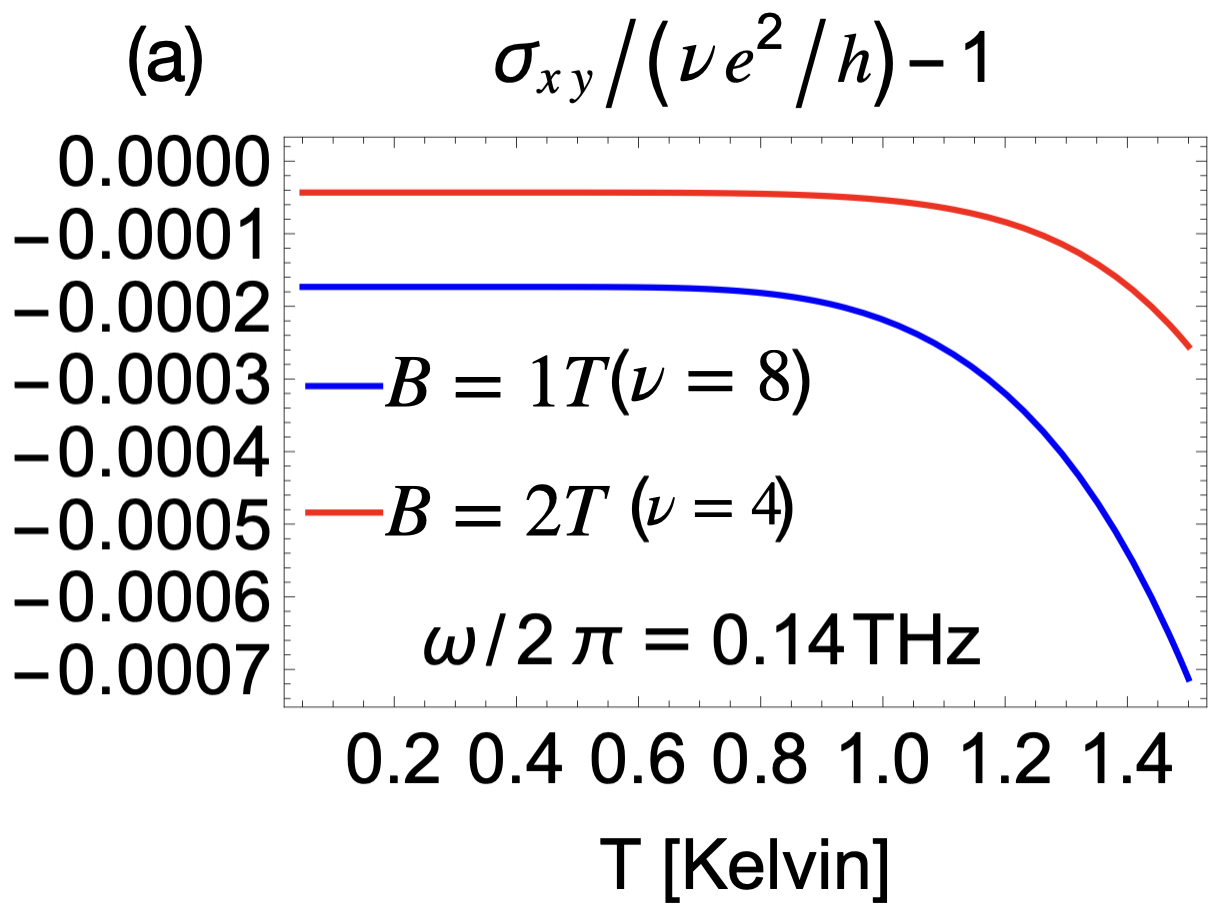}
       \end{subfigure}
    \hspace{-0.05cm}
     \begin{subfigure}[b]{0.35\textwidth}
         \centering
         \includegraphics[width=\columnwidth]{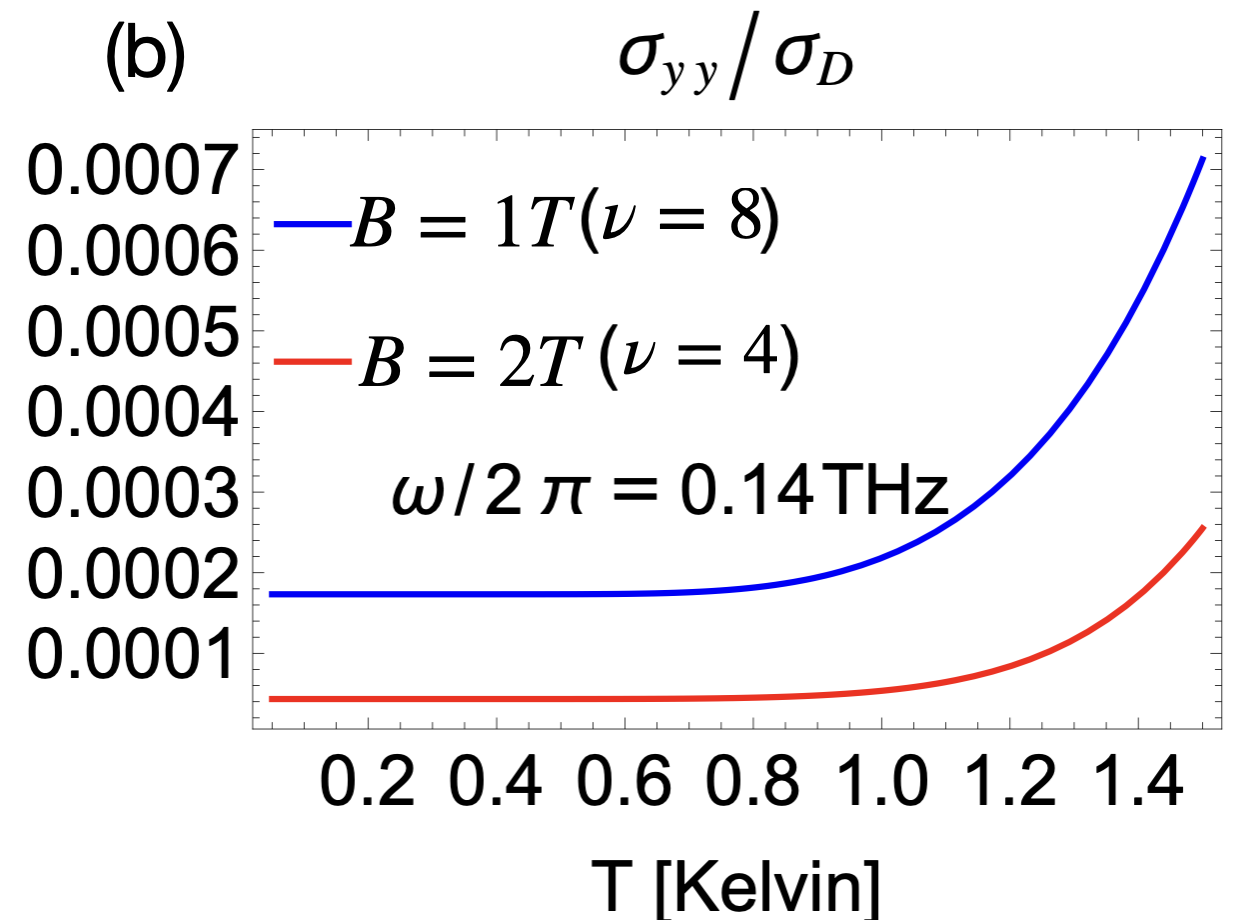}
 \end{subfigure}
    \caption{Finite temperature quantum Hall transport in the cavity. The cavity frequency is $\omega=2\pi \times 0.14 \textrm{THz}$ and 2d electron density $n_{\textrm{2d}}=2\times 10^{11}\textrm{cm}^{-2}$. The polariton broadening is $\delta=2\pi\times 5\times 10^{-3}\textrm{THz}$ as in Fig.\ref{Lifetime} in the main text. a) Deviations of the Hall conductance from the expected quantized values for two different plateaus $\nu=8,4$ corresponding to $B=1,2T$. b) Thermal activation of the longitudinal conductance $\sigma_{yy}$ normalized by the Drude DC conductivity $\sigma_D=e^2n_{\textrm{2d}}/m\delta$. For low temperatures $T<0.4K$ the finite temperature transport reproduces the $T=0$ transport in Fig.\ref{Lifetime} for $\omega=2\pi\times 0.14\textrm{THz}$ presented in the main text. }
       \label{supp_transport}
\end{figure}

For completeness and as supporting computation, in Fig.~\ref{supp_transport} we provide the finite temperature transport of the polariton system for the experimentally reported parameters in Ref.~\cite{FaistCavityHall}, $\omega=2\pi\times 0.14 \textrm{THz}$ and $n_{\textrm{2d}}=2\times 10^{11}\textrm{cm}^{-2}$, and broadening $\delta=2\pi\times 5\times 10^{-3}\textrm{THz}$ which we used in Fig.~\ref{Lifetime} in the main text. In Fig.~\ref{supp_transport}(a) we show the deviation of the Hall conductance from the topologically expected quantized values for two different plateaus $\nu=8,4$ ( $B=1,2T$). In Fig.~\ref{supp_transport}(b) we show the thermal behavior of the longitudinal conductance $\sigma_{yy}$. We observe the exponential thermal activation as expected. In the low temperature regime, $T<0.4\textrm{K}$, we see that the modifications of quantum Hall transport are consistent with the $T=0$ results show in Fig.~\ref{Lifetime} in the main text. Being more precise, the cavity-induced transport deviations at $T<0.4K$ and for $B=1T (\nu=8)$ are $\sim 2\times 10^{-4}$ and for $B=2T (\nu=4)$ are $\sim 5\times 10^{-5}$. These values agree with the $T=0$ transport for $\omega=2\pi\times0.14 \textrm{THz}$ shown in Fig.~\ref{Lifetime} in the main text.    

\subsection{Zero Temperature Transport}

Having derived the general formula for the current correlator $\chi_{ab}(w)$ at finite temperature, we will focus now at the transport properties at zero temperature, $T=0$, where the topological protection and the quantization of the quantum Hall conductance are expected from the Thouless argument, as long as the system is gapped~\cite{Thouless}. At $T=0$ the ground state of the system is for $n_+=n_-=0$ and only the thermal prefactors corresponding to the ground state $e^{-\beta E_{00}}$ contribute to transport. 
\begin{equation}
    \chi_{ab}(w)=\sum_{m_+,m_-} \frac{\langle 00|J_{a}|m_+m_-\rangle\langle m_+m_-|J_{b}|00\rangle}{w+(E_{00}-E_{m_+m_-})/\hbar +\textrm{i}\delta} - (00 \leftrightarrow m_+m_-)
\end{equation}
Furthermore, the current operators are linear in the polaritonic annihilation and creation operators and thus allow only for single-polariton transitions to occur, which implies that inthe denominator of the response function only single polariton energies show up $\Omega_{\pm}$. Finally, using the formulas for the matrix representation of the components of the current operator we find the following analytically exact expressions for the transverse $\chi_{xy}$ and longitudinal $\chi_{yy}$ response functions
\begin{eqnarray}
 \chi_{xy}(w)&=&\frac{Ne^3B}{(1+\Lambda^2)m^2_e}\Big[ \Lambda(\Lambda+\eta)\frac{\textrm{i}}{2}\left(\frac{1}{w+\Omega_-+\textrm{i}\delta}+\frac{1}{w-\Omega_-+\textrm{i}\delta}\right)
 +(1-\eta\Lambda)\frac{\textrm{i}}{2}\left(\frac{1}{w+\Omega_++\textrm{i}\delta}+\frac{1}{w-\Omega_++\textrm{i}\delta}\right)\Big],\nonumber\\
\chi_{yy}(w)&=&-\frac{Ne^2}{(1+\Lambda^2)m}\left[\frac{\Omega_+}{2}\left(\frac{1}{w+\Omega_++\textrm{i}\delta}-\frac{1}{w-\Omega_++\textrm{i}\delta}\right)+\frac{\Lambda^2\Omega_-}{2}\left(\frac{1}{w+\Omega_-+\textrm{i}\delta}-\frac{1}{w-\Omega_-+\textrm{i}\delta}\right) \right].
\end{eqnarray}
With the above results and using the Kubo formula, the expressions for the optical and the CD conductivities can be straightforwardly obtained.

\end{document}